# Amplification of electromagnetic waves excited by a chain of propagating magnetic vortices in YBa$_2$Cu$_3$O$_{7-\delta}$ Josephson-junction arrays at 77K and above

Boris Chesca[1*], Daniel John[1], Christopher J. Mellor[2]

[1]Department of Physics, Loughborough University

Loughborough, United Kingdom

[2]School of Physics and Astronomy, Nottingham University

Nottingham, United Kingdom

**Corresponding author:**

Dr. Boris Chesca

Physics Department

Loughborough University

Loughborough,

UK

Email: B.Chesca@lboro.ac.uk

Phone: +44(0)1509223985

**Abstract**

When a soliton propagates in a discrete lattice it excites small-amplitude linear waves in its wake. In a dc current-biased Josephson-junction (JJ) array these manifest as electromagnetic (*EM*) waves excited by a (magnetic field induced) chain of propagating magnetic vortices. When the vortex velocity and the phase velocity of one of the excited *EM* waves match, phase-locking occurs. This produces resonant steps in the current-voltage characteristics where amplification of *EM* radiation occurs. We report the first observation of phase-locking-induced amplification of *EM* radiation at 77K and above in *JJ* arrays made of high temperature superconductors.

Soliton propagation in a discrete lattice is used to model the dynamics of many systems [1, 2] such as magnetic and ferroelectric domain walls, dislocations, optical fibres, biological neurons, or Josephson-junction arrays (in both their realizations: with superconductors [3-10] or atomic Bose-Einstein condensates [11]). In particular, when a magnetic field, *B*, is applied perpendicular to a dc current-biased planar one-dimensional (1D) Josephson–junction (JJ) array, magnetic vortices will enter the array in a form of Josephson vortices. The bias current, *I*, flowing across the array produces a Lorentz force, $F_L$, which drives the Josephson vortices unidirectionally, forming a lattice of vortices moving with a speed, *v* (see Fig. 1). As for other cases where solitons propagate in discrete lattices, this is accompanied by an emission of small amplitude linear waves that propagate along the array [3, 4]. In magnetic systems these waves take the form of phonons excited by the propagating domain walls. In a dc current-biased JJ array these manifest as electromagnetic (*EM*) waves excited by a (magnetic field induced) chain of propagating magnetic vortices. When the vortex spacing is commensurate with the wavelength of emitted waves, resonant modes occur. This can be viewed as the phase-locking condition of the vortex velocity and the phase velocity of one of the self-induced modes. The experimental signature of such phase locking between a train of propagating vortices and their induced *EM* radiation in a JJ array is a series of flux-flow resonances in the current-voltage characteristics (IVCs) [3, 4]. On a resonant current step, moving vortices couple to their induced linear waves. Further increases of the current do not lead to further linear increases in the vortices velocity, because the energy is consumed in amplifying the linear waves. Such resonances have been observed by many groups, below 9K, in JJ arrays made of low temperature superconductors (LTS) [4-7]. The associated amplified *EM* radiation has been measured too, proving that LTS JJ arrays are suitable candidates as *B*-tunable microwave oscillators [8, 9]. After the discovery of high temperature superconductors (HTS) in 1986, it was anticipated that similar *B*-tunable microwave oscillators could be made of HTS JJ arrays operating at liquid nitrogen temperature. However the highest previously reported temperature for observing stable flux-flow resonances in HTS JJ arrays is about 40K, recently achieved in $Bi_2Sr_2CaCu_2O_{8+x}$ mesa structures based on a 1D array of intrinsic JJs [10]. Here we report the first observation of multiple flux-flow resonances at 77K and above in a parallel array of 440 $YBa_2Cu_3O_{7-\delta}$ bicrystal grain boundary JJs. This strongly suggests that phase locking-induced enhanced *EM* radiation has been excited in HTS JJ arrays operating at liquid nitrogen temperature and above.

Dc current biased JJ arrays in the presence of an applied B field are governed by the discrete sine-Gordon equations,

$$\frac{\ddot{\phi}_i}{\omega_p^2} + \frac{1}{\beta_c \omega_p}\dot{\phi}_i + \sin\phi_i = \frac{1}{\beta_L}\left(\phi_{i+1} - 2\phi_i + \phi_{i-1}\right) + \frac{I}{I_{cJ}} \qquad (1)$$

with $i=1, \ldots, N$, N being the number of junctions, $\phi_i$ the phase difference across the ith junction, $\beta_c = 2\pi R_N^2 C I_{cJ}/\Phi_0$ the Stewart-McCumber junction parameter, $\omega_p=(2\pi I_{cJ}/C)^{1/2}$ the junction plasma frequency, $R_N$ is the JJ normal resistance, and $C$ the JJ capacitance. The overdots denote differentiation with respect to time. The discreteness parameter $\beta_L$ is defined as $\beta_L = 2\pi L I_{cJ}/\Phi_0$ where $\Phi_0$ is the flux quantum, $I_{cJ}$ is the maximum Josephson critical current of each junction and $L$ is the inductance of the superconducting loop between $JJ_i$ and $JJ_{i+1}$. Because the system is discrete, we expect its properties to be periodic in the applied flux $\Phi_{ex}$ (proportional to B) with period $\Phi_0$. Eqs (1) have been investigated theoretically and subsequently, families of IVCs for different values of $\Phi_{ex}$ have been calculated [3-7]. Multiple $m$ ($m=1, 2, 3, 4,\ldots$) flux flow resonances have been identified on the IVCs. An approximate analytical expression has been found [5] for their voltage location on $\Phi_{ex}$ for the case when resonances are not too strong:

$$\frac{V}{I_{cJ} R_N} = \frac{2}{m\sqrt{\beta_L}} \sin\left( m\pi \frac{\Phi_{ex}}{\Phi_0} \right) \tag{2}$$

Within each period (0, $\Phi_0$) Eq. (2) is only valid for applied fluxes larger than a cut-off value $\Phi_{c1}$ known as the minimum applied flux density required for vortices to enter the array.

The physics behind the oscillations of flux flow resonances with $\Phi_{ex}$ as given by Eq. (2) is well understood [6, 7] as summarized below for $m=1$. When increasing the applied flux starting from 0 to $\Phi_0/2$ per cell the voltage location of the first ($m=1$) flux flow resonance increases since an increasing number of fluxons (proportional to B) enter the array and perform a unidirectional motion. We call this behaviour the fluxon-motion mode. However, at a magnetic flux close to $\Phi_0/2$ the flux flow voltage saturates. When increasing the flux further from $\Phi_0/2$ to $\Phi_0$ the flux flow voltage decreases. This happens because for fluxes larger than $\Phi_0/2$ a fixed number of fluxons, commensurate with the lattice remains pinned and the most energetically favourable dynamical state is a motion of vacancies in this fluxon chain instead of a motion of all the fluxons themselves. We call this behaviour the fluxon-vacancies-motion mode. Experimentally only the first two resonances ($m=1$ and $m=2$) have been investigated in detail by earlier studies [4-7]. In this report we investigated experimentally resonances up to $m=4$ and we find several novel features regarding their coexistence and mutual interaction. Thus, from Eq. (2) it is expected that as we increase the applied flux from 0, there should be a mode-crossing event between $m=2$ and $m=3$ modes twice every period of $\Phi_0$ and between $m=3$ and $m=4$ modes four times every period of $\Phi_0$. We have observed such mode-crossing events for the first time and found that each time a mode-crossing event occur both modes involved do not interfere in any way (destructively or constructively) with each other but continue to manifest totally independent from one another.

To clarify the physical origin of the resonances, it is helpful to recall a mechanical analogue of the system. Eqs. (1) may be viewed as the equations of motion for a chain of N identical pendulums, each of which is viscously damped and free to move transverse to the axis of the chain, driven by a constant torque, and coupled to its nearest neighbours by torsional springs [4]. A vortex corresponds to a kink (soliton) propagating along the chain. In this configuration, a given pendulum hangs almost straight down for much of the time, but when the kink passes by, the pendulum overturns rapidly and oscillates for the period between passing kinks. These oscillations are the analogue of the *EM* radiation excited by the kink. A resonance occurs if the pendulum oscillates precisely an integer number of times ($m$) between successive passages of the kink; or equivalently, when the vortex lattice moves at the same velocity as one of the modes $m$. The possible oscillation frequencies are the lattice eigenfrequencies of small oscillations about the kink. Each junction (pendulum) has an identical behaviour except for a constant shift in time. This suggests that the solutions are well approximated by travelling waves. Near both ends of the array, vortex reflections from the ends change this picture and destructively interfere with the flux flow resonances [5]. To minimize this effect we designed our JJ array as a very long ($N$=440) and asymmetric array consisting of 22 identical sets of 20 junctions (pendulums) each (see Fig. 1a). As we move from the right edge to the left edge of the array in Fig. 1, within each set of 20 junctions (pendulums), the loop inductance $L_i$ (the coupling strength of torsional strings $K_i$) between $JJ_i$ and $JJ_{i+1}$ (pendulums $i$ and $i+1$) decreases monotonically from 1 to 0.7 in normalized units. This inductance asymmetry along the array is small enough to preserve the validity of Eq. (1). The asymmetry, however, is sufficiently strong to ensure that the vortex–flow is enhanced when vortices are moving to the left (in the direction of decreasing loop inductance within each set of 20 consecutive JJ) and is suppressed when they are moving to the right. In addition this asymmetry, ensures that the reflected vortices do not propagate backwards deep into the array to destructively interfere with the incoming train of vortices. This protects the flow of vortices in one direction (to the left in Fig.1) and the associated EM radiation amplification. To understand the asymmetry in the vortex flow again it is helpful to recall the mechanical analogue of a chain of variable harmonic coupling $K_i$ pendulums: it is easier for a kink to propagate in the direction of decreasing $K_i$ than in the opposite direction. A similar asymmetric $K_i$-induced ratchet potential that lacks inversion symmetry producing such a preferential directional motion for fluxons in JJ arrays has been previously investigated theoretically in [12]. As a result of this asymmetry in the flux-flow direction, at all temperatures investigated (77-92)K the flux-flow resonances observed were more pronounced for positive bias currents (the direction of $I$ in Fig.1) than for negative ones. It was even possible for vortices to propagate in the low damping direction only. Thus, in the experiments conducted at 89K the flux-flow resonances are observed for positive bias currents only (see Fig. 2).

The JJ arrays were fabricated by depositing high quality epitaxial, 100nm thick *c*-axis oriented YBa$_2$Cu$_3$O$_{7-\delta}$ (YBCO) films on 10x10 mm$^2$, 24° symmetric [001] tilt SrTiO$_3$ bicrystals by pulsed laser deposition [13]. A 200 nm thick Au layer was deposited in situ on top of the YBCO film to facilitate fabrication of high quality electrical contacts for electric transport measurements. The films, with a critical temperature of $T_c$ of 92K, were subsequently patterned by optical lithography and etched by an Ar ion beam to form a parallel array of 22 x 20 *JJs*. All 440 JJs are 3 μm wide. The junctions are separated by superconducting loops of identical width of 3 μm but variable length. Thus, within each set of 20 *JJs* the length of loops varies logarithmically from 13 μm to 8 μm. Consequently, the $\beta_{Li} = 2\pi L_i I_{cJ}/\Phi_0$ is also changing monotonically within each set of 19 superconducting loops (from 1 to 0.7 in normalized units). An optical micrograph of a small part of such an array (containing 42 JJs including 2 full sets of 20 *JJs* each) is shown in Fig. 1a. *I* is applied symmetrically via the central top and bottom electrodes and V is measured across the array. A magnetic field, **B**, is applied perpendicular to the planar array's structure via a control current $I_{ctrl}$. Consequently, an external magnetic flux $\Phi_{ex}$ is coupled into the array. We have fabricated two such devices and both showed a similar behaviour. Each JJ can be modelled as a capacitively and resistively shunted Josephson junction so that the equivalent circuit of a JJ array is a parallel array of such elements connected via superconducting inductances $L_i$ (see Fig. 1b).

Families of *IVC's* were measured by a 4 point-contact method at various temperatures between 77K and 92K and for different values of the control current $I_{ctrl}$ in the range (-8mA, 8mA). $I_{ctrl}$ was changed in small steps of 15μA. This allows a scan over $\Phi_{ex}$ that spreads over more than 8 periods in Eqs. (1) and (2) at a given temperature with about 130 IVC's per period. For all temperatures measured the current step corresponding to *m*=1 resonance had an amplitude $I_{step}$ about the same as the array's maximum critical current $I_c$. A family of 30 consecutive *IVC's* measured at 89K are plotted in Fig. 2. From such families of *IVC's* scanned over $I_{ctrl}$ (or equivalently, $\Phi_{ex}$) $I_c(\Phi_{ex})$ for both positive and negative currents could be constructed (see inset of Fig. 2). At temperatures close to $T_c$ the discreteness parameter $\beta_{Li}$ is negligibly small. In this limit Eqs. (1) correspond to the continuum case, i.e., a single long JJ, where a single flux-flow resonance should appear [14]. Also $I_c$ versus $\Phi_{ex}$ should have a periodic pattern consisting of a series of maxima similar to a diffraction pattern of an optical multiple slit grating [15]. The flux periodicity corresponds to one additional flux quantum $\Phi_0$ in each loop. All these features are observed in the experiments carried out at 89K (see the $I_c(\Phi_{ex})$ dependence for negative bias currents in the inset of Fig. 2). A single large flux-flow resonance is clearly visible for positive bias currents only (see Fig. 2). Its presence strongly affects the $I_c(\Phi_{ex})$ dependence for positive values which now consists of a periodic two peak structure instead. Such a current step resonance, corresponding to *m*=1, is tunable by varying **B**, i.e., its voltage location strongly depends on $\Phi_{ex}$ (shown in red in the inset of Fig. 2) and, as expected, it has the same periodicity as $I_c(\Phi_{ex})$. The resonance reaches its maximum voltage location at flux values that correspond to a minima in $I_c(\Phi_{ex})$, as observed previously in LTS JJ arrays [6-7]. No such resonance is observed for negative bias currents due to the inductance asymmetry along the array, as explained earlier. At lower temperatures (77-84)K $\beta_{Li}$ increases due to an increase in $I_{cJ}$ and, as expected, multiple flux-flow resonances are observed corresponding to *m*=1, 2, 3 and 4 (see Figs. 3 and 4). Again, due to inductance asymmetry along the array, these resonances are more pronounced for positive current biases. Figs. 3a and 3b show families of 20 and 16, respectively, consecutive IVC's at 84 K where multiple current step resonances are clearly visible. In Fig. 3a, as we increase $\Phi_{ex}$, there is mode-crossing between *m*=2 resonance moving to the left and *m*=3 resonance moving to the right (we refer to this event as to a (2, 3) mode-crossing). The inset of Fig. 3a shows the derivatives *dI/dV* of 7 consecutive IVC's (5-12) just before, during, and after a (3, 4) mode crossing. Both m=3 and m=4 resonances, seen as peaks in the *dI/dV* versus V characteristics, are clearly visible just before and after the mode-crossing. At the mode-crossing point both peaks merge as one. Since both *I* and *V* are the same at the mode-crossing point it means that all 440 JJs oscillate (synchronized with a constant shift in time) 3 times between successive passages of a chain of vortices (*m*=3 resonance) and, at the same time, they all oscillate 4 times between successive passages of another chain of vortices in the same direction (*m*=4 resonance). Similar coexistence of two resonance modes have been observed in the (2, 3) mode-crossing events as well (see inset of Fig. 4). It follows that both resonance modes involved in a mode-crossing survive such events suggesting that both modes do not interfere destructively with each other but coexist instead. In Fig. 3b, resonance *m*=2 has a reduced *B*-field tunability in comparison with all the other 3 resonances (*m*=1, 3, and 4). Multiple flux-flow resonances up to *m*=4 have been observed at 81K and 77K as well, revealing a qualitatively very similar behaviour as summarized in Figs. 3a and 3b for *T*=84K. With decreasing temperature these resonances become steeper and larger. An example is shown in Fig. 4 where the *m*=2 current resonance at 77K reaches a height of 0.3 mA, about the same as the value of the array's critical current.

The voltage location of resonances versus $\Phi_{ex}$ at 84K is shown in Fig. 5. The qualitative agreement with Eq. 2 is remarkable considering that no fitting parameter has been used in this comparison. However, it is important to stress that Eq. 2 is oversimplified as neither the vortex depinning term [16], nor the dissipation (which might be significant considering the measured IVCs are non-hysteretic), nor the variable $\beta_{Li}$ along the array have been considered when deriving Eq. 2. This explains the quantitative discrepancies between theory and experiments in Fig. 5. The resonances disappear at a cut-off value (called $\Phi_{c1}$ for m=1 in Fig. 5) which, in general, is different for every mode. It has been shown previously that, precisely around $\Phi_{c1}$, the system performs as a Josephson vortex-flow transistor with very high current gains [17].

Fourier spectral analysis of the flux-flow voltages reveal that on the step *m*, the first *m* harmonics are dominant, and the higher harmonics have rapidly decaying amplitudes [5]. Therefore, for possible applications as *B*-field tunable oscillators, operation on *m*>1 steps is not desirable since the ac power will be distributed among the modes, instead of being concentrated in one mode as for *m*=1. The phase locking-induced amplification coefficient of the *EM* waves is periodic in the applied flux $\Phi_{ex}$ with period $\Phi_0$ and reaches its maximum at $\Phi_{ex}=\Phi_0/2$ for the *m*=1 resonance and at $\Phi_{ex}=\Phi_{c1}$ for *m*=2 resonance [5]. With our device operated at 84K on the *m*=1 step, one can estimate a maximum power radiation $P=0.5(I \times I_{step}/I_c)^2 R_N/N$ of about 20 nW at $\Phi_{ex}=\Phi_0/2$ and a *B*-field tunability within the frequency range (1-9) GHz. At 77K the device performance improves considerably providing a *B*-field tunable power of about 0.1 µW within the range (1.5-25) GHz. As shown in [8, 9], by inductively coupling multiple *M* rows of 1D LTS JJ array the power of emitted *EM* radiation can be further increased. The increase is most significant (proportional to *M*) if operating the device at *m*=2 resonance [9]. A similar concept can be used for the implementation of 2D (*M*×*N*) HTS JJ arrays using our design.

Phase-locking in a parallel array of 440 YBa$_2$Cu$_3$O$_{7-\delta}$ *JJs* between a chain of propagating vortices and their induced electromagnetic waves has been demonstrated experimentally in the temperature range 77K-89K. We showed that phase-locking produces multiple (*m*=1, 2, 3, and 4) flux-flow resonances on the measured *IVCs* where amplification of vortex propagation-induced *EM* radiation occurs. Similar resonant amplification of phonons has been observed in propagating magnetic domain walls [18]. Remarkably, these *B*-field tunable resonances remain stable over a wide range of *B* and *T* without any magnetic or electric field shielding suggesting the strong coherent behaviour of hundreds of junctions at liquid nitrogen temperature and above. Thus, the design approach reported here shows great promise as a route to realizing high performance B-field tunable HTS microwave oscillators based on flux flow resonances. Our results also show that HTS *JJ* arrays are ideal candidates for studying non-linear wave propagation in discrete lattices and their related phase-locking phenomena.

**Acknowledgments**

The authors would like to thank Dominic Walliman and Jas Chauhan for their technical support.

**References**

[1] J. F. Currie, S. E. Trullinger, A. R. Bishop, and J. A. Krumhansl, *Phys. Rev. B* **15**, 5567-5580 (1977).

[2] M. Payrard and M. D. Kruskal, *Physica D* **14**, 88-102 (1984).

[3] A. V. Ustinov, M. Cirillo, and B. A. Malomed, *Phys. Rev. B* **47**, 8357-8360 (1993).

[4] H. S. J. Van der Zant, T. P. Orlando, S. Watanabe, and S. H. Strogatz, *Phys. Rev. Lett.* **74**, 172-177 (1994).

[5] A. E. Duwel, et al., *J. Appl. Phys*. **82**, 4661-4668 (1997).

[6] A. V. Ustinov, et al., *Phys. Rev. B* **51**, 3081-3091 (1995).

[7] M. Cirillo, et al., *Phys. Lett. A* **183**, 383-389 (1993).

[8] P. A. A. Booi, and S. P. Benz, *Appl. Phys. Lett.* **64**, 2163-2165 (1994).

[9] A. E. Duwel, T. P. Orlando, S. Watanabe, S., H. S. J.Van der Zant, *IEEE Trans. Appl. Supercond.* **7**, 2897-2900(1997).

[10] S. O. Katterwe and V. M. Krasnov, *Phys. Rev. B* **84**, 214519 (2011).

[11] F. S. Cataliotti, et al., Science, 293, 843 (2001).

[12] F. Falo, P. J. Martinez, J.J. Mazo, and S. Cilla, Europhys. Lett., 45, 700 (1999).

[13] The films were deposited by TESA GmbH.

[14] R. E. Eck, D.J. Scalapino, and B. N. Taylor, *Phys. Rev. Lett.* **13**, 15-18 (1964)

[15] J. H. Miller, G. H. Gunaratne, J. Huang, and T. D. Golding, *Appl. Phys. Lett*. **59**, 3330-3332 (1991).

[16] Z. Zheng, B. Hu, and G. Hu, *Phys. Rev. B* **58**, p5453 (1998).

[17] B. Chesca, D. John, M. Kemp, J. Brown, and C.J. Mellor, *Appl. Phys. Lett.* **103**, 092601 (2013).

[18] L. Thomas, et al., *Science* **315**, 1553-1556 (2007).

**Figure captions**

Fig.1. (a) Optical micrograph showing part of a parallel array of 440 $YBa_2Cu_3O_{7-\delta}$ bicrystal grain boundary JJs consisting of 22 identical sets of 20 identical JJs. The bicrystal grain boundary (GB) is shown with dotted line. The Josephson junctions are formed across the GB (seen as grey colour areas crossing the GB). Shown is a small central part of the array consisting of 42 JJs including 2 full sets of 20 JJs. Within each set the JJs are connected via superconducting loops of identical width but monotonically decreasing length. This inductance asymmetry along the array provides an enhanced vortex flow to the left relative to vortex propagation to the right. (b) Schematic view of the equivalent circuit of a JJ inductively connected to its neighbor JJ.

Fig.2. A family of 30 consecutive *IVC*s recorded at 89K for different values of $I_{ctrl}$. $I_{ctrl}$ is changed in steps of 15 µA. The horizontal arrow shows the resonance's direction of motion in the *IV* plane when increasing B. The family is shown in 3 sets of 10 *IVC*'s each. For clarity *IVC*s 11 to 20 (21 to 30) have been shifted vertically by 0.1 mA (0.2 mA). In blue (red) is the first (last) *IVC* within each set. The inset shows $I_c(\Phi_{ex})$ for both positive and negative currents. In red is the flux-flow resonance voltage (normalized to the $I_{cJ}R_N$ product of 7.5 µV at 89K) versus $\Phi_{ex}$.

Fig.3. Families of *IVC*s for different values of $I_{ctrl}$ at 84K. The horizontal arrows show the directions of motion of the resonances in the *IV* plane when increasing *B*. (a) A family of 20 consecutive *IVC*s shown in 2 separate sets of 10 *IVC*s each. In blue (red) is the first (last) *IVC* within each set. For clarity *IVC*s 11 to 20 have been shifted vertically by 0.2 mA. Inset shows the derivatives *dI/dV* of *IVC*s 5-12 before and after a (3, 4) mode-crossing event. (b) A family

of 16 consecutive *IVC*s. Inset shows the derivatives *dI/dV* of *IVC*s 1-8 where all 4 peaks corresponding to $m=1, 2, 3, 4$ modes are clearly visible .

Fig.4. Families of 16 *IVC*s for different values of $I_{ctrl}$ at 77K. The horizontal arrows show the directions of motion of resonant modes 2 and 3 in the *IV* plane when increasing *B*. Inset shows the derivatives *dI/dV* of all odd *IVC*s between 1-16 before and after a (2, 3) mode-crossing event.

Fig.5 Flux-flow resonance voltages (normalized to the $I_{cJ}R_N$ product of 18.5 µV at 84K) versus $\Phi_{ex}$. Full lines are given by Eq. (2) with $m=1, 2, 3$, and 4, while circles are experimental data. Vertical dotted lines show the ranges of the *B* field where the families of *IVC*s plotted in Fig. 3a and Fig. 3b were measured.

Figure 1

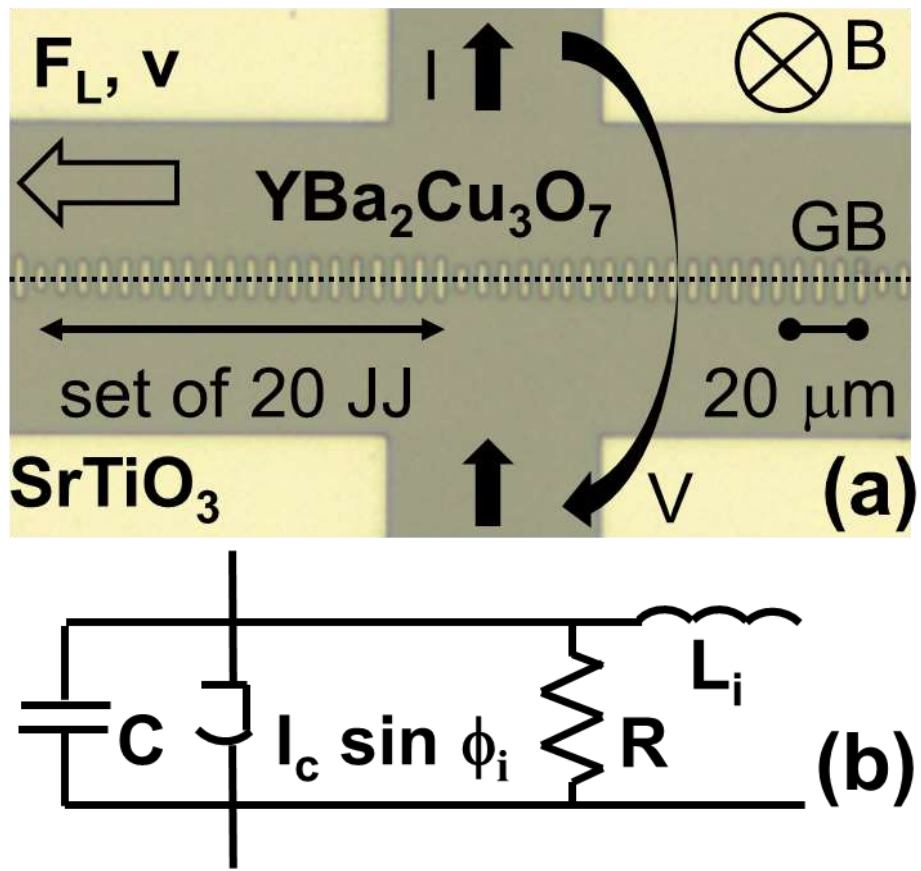

Figure 2

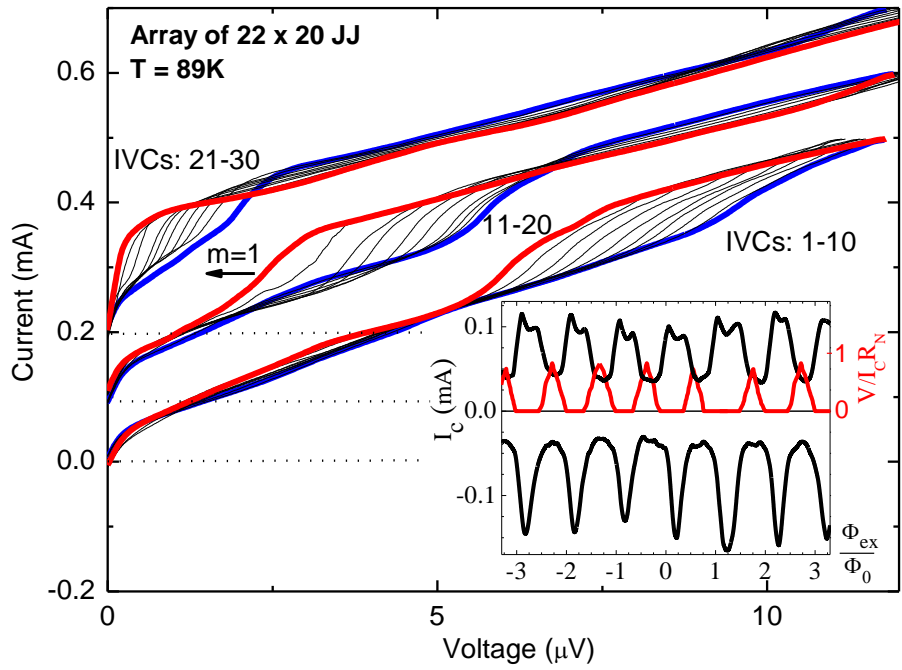

Figure 3

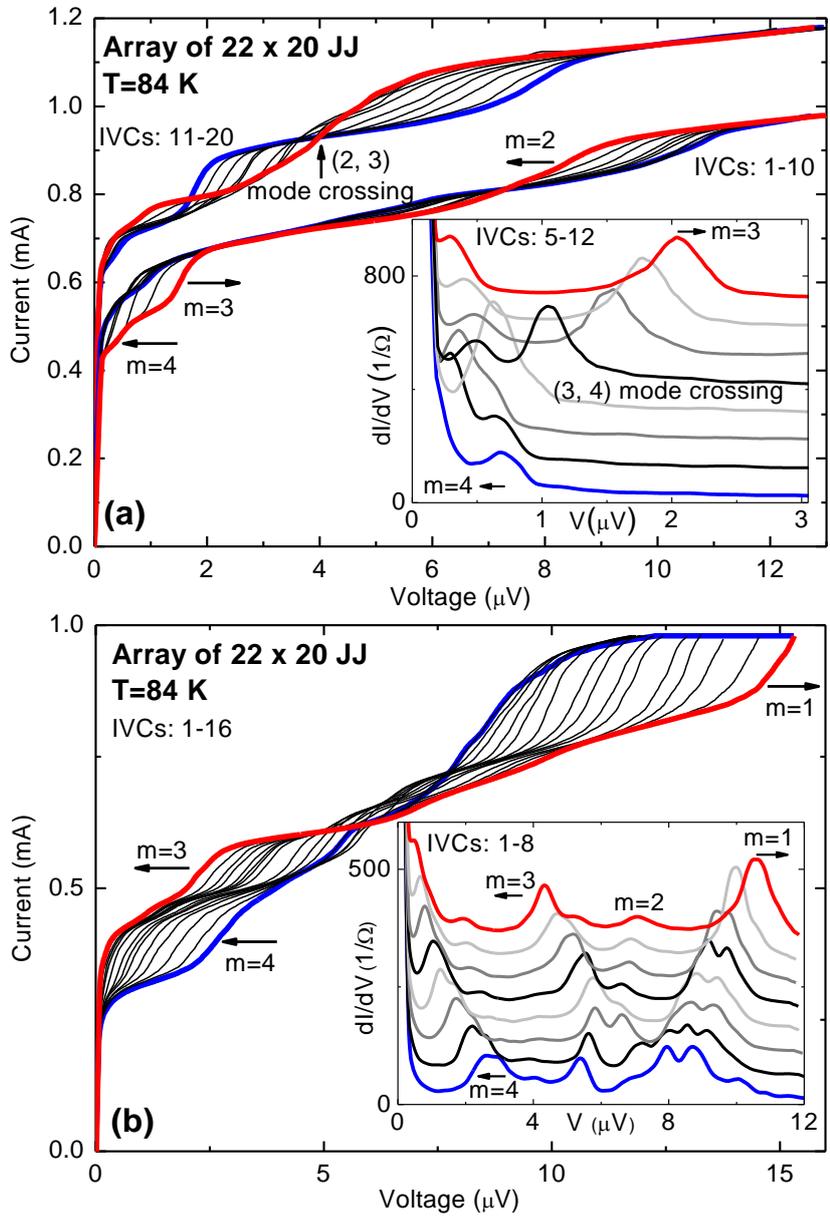

Figure 4

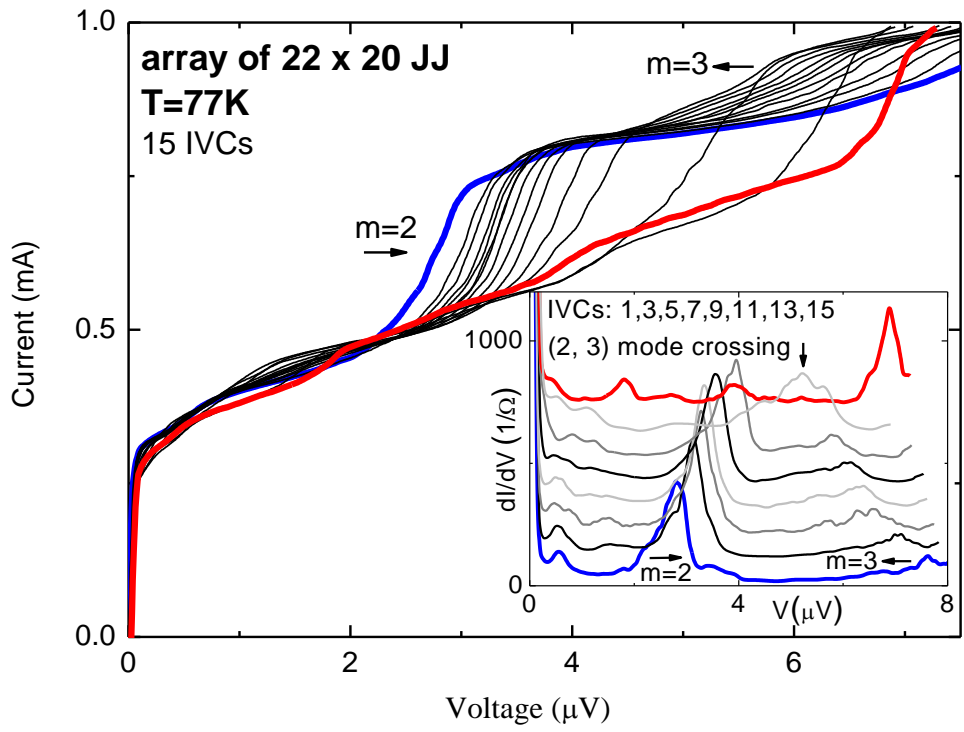

Figure 5

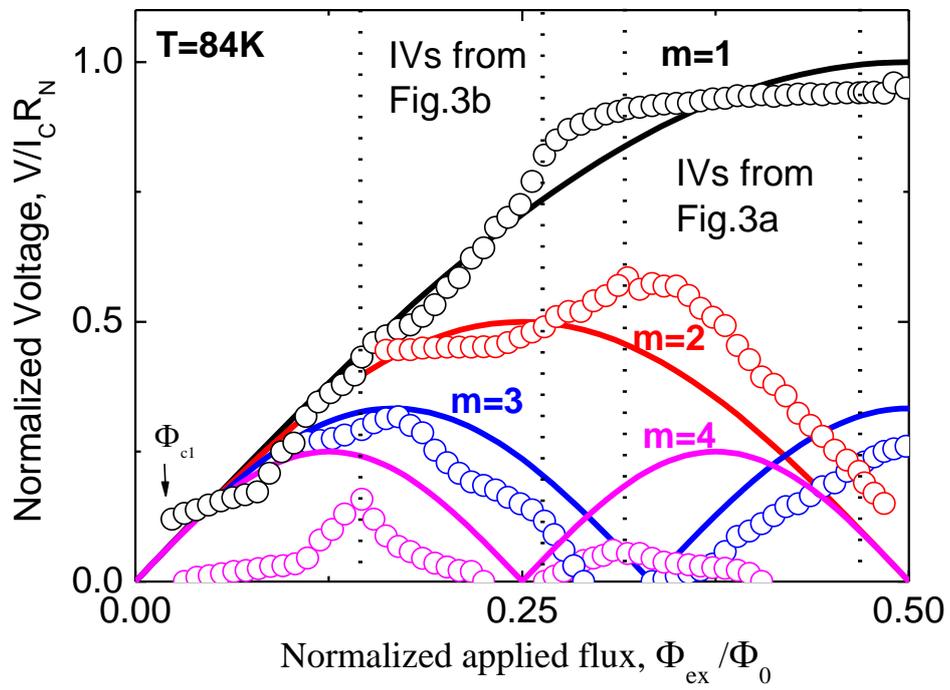

19